\documentclass[nofootinbib,prd,12pt,superscriptaddress]{revtex4}%
\usepackage{amsmath}
\usepackage{amsfonts}
\usepackage{amssymb}
\usepackage{graphicx}%
\usepackage{amsmath,amssymb}
\usepackage{mathrsfs}
\usepackage{graphicx}
\usepackage{color}
\usepackage{subfigure}
\usepackage{fancyhdr}
\usepackage{multirow}
\usepackage{float}
\usepackage{epsfig}
\usepackage{amsfonts}
\usepackage{bm}

\def\be{\begin{equation}}
\def\ee{\end{equation}}
\def\beq{\begin{eqnarray}}
\def\eeq{\end{eqnarray}}
\def\ba{\begin{eqnarray}}
\def\ea{\end{eqnarray}}

\begin{document}

\title{High-Temperature Modifications of Charged Casimir Wormholes}

\author{Phongpichit Channuie}
\email{phongpichit.ch@mail.wu.ac.th}
\affiliation{School of Science, Walailak University, Nakhon Si Thammarat, 80160, Thailand}
\affiliation{College of Graduate Studies, Walailak University, Nakhon Si Thammarat, 80160, Thailand}

\date{\today}

\begin{abstract}

In this work, we extend the investigation of the consequences of thermal fluctuations on the Casimir effect within the context of a traversable wormhole, recently proposed by Garattini \& Faizal, arXiv:2403.15174 [gr-qc], subject to charge contributions. Specifically, we focus on scenarios where the plates exhibit both constant and radial variations. In our analysis, we initially concentrate on the high temperature approximation, considering solely the influence of charge on the thermal Casimir wormholes. Additionally, upon incorporating Generalized Uncertainty Principle (GUP) corrections to the Casimir energy, we obtain a new class of wormhole solutions. Notably, we establish that the flare-out condition remains consistently satisfied. Intriguingly, our findings reveal that both the charge and GUP contributions serve to further enlarge the throat's size in the radial variation.

\end{abstract}

\maketitle


\section{Introduction}
The Casimir effect is based on the assumption of perfectly conducting surfaces. However, real-world plates are never perfectly conducting; they exhibit a complex permittivity, varying with frequency depending on the material. This consideration alters the original Casimir force at zero temperature, transforming it into a finite-temperature thermal effect. This thermal effect incorporates both thermal and quantum fluctuations, as described by the Lifshitz theory, which accounts for correlated fluctuating charges and currents in the plates \cite{Sabisky1973}. This adaptation of the conventional Casimir effect to finite temperature is termed the thermal Casimir effect \cite{Canaguier-Durand:2010hwu,Marino:2014rfa}. It has been shown that at separations beyond a critical value, the force generated by thermal fluctuations surpasses that of zero-temperature quantum fluctuation \cite{Sushkov:2010cv,Bostrom:2000zza}. This suggests the potential significance of thermal effects in scaling up the size of traversable wormholes (TWs). Very recently, the consequences of thermal fluctuations to the Casimir effect on a traversable wormhole have been investigated \cite{Garattini:2024jkr}. The authors illustrated how thermal fluctuations alter the throat of the wormhole in both high and low temperature regimes. Furthermore, they elucidated the impact of finite temperature on the size of a wormhole. In this work, we investigate new exact and analytic solutions of the Einstein–Maxwell field equations describing Casimir wormholes including the high-temperature contribution to the Casimir energy source for a traversable wormhole with the plates radially varying and discuss the effect of the Generalized Uncertainty Principle (GUP). Notice that the electromagnetic field combined with the Casimir source has been used to examine the impacts of such an electrovacuum on a traversable wormhole \cite{Samart:2021tvl,Garattini:2023qyo}.

The main purpose of this work is to extend the analysis present in Ref.\cite{Garattini:2024jkr} for the thermal Casimir wormholes by incorporating  the electric charge underlying the GUP. This work is organized as follows: In Section (\ref{sec2}), we formulate the Einstein–Maxwell field equations describing Casimir wormholes with charge. We next consider the thermal Casimir wormholes employing energy densities featuring the thermal modifications to the chared Casimir wormholes and solve to obtian the wormholes solutions in Section (\ref{sec3}). The results also include the GUP effects. We conclude of findings in the last section.

\section{Setup}\label{sec2}
Our aim is to model a wormhole solution and typically we consider an anisotropic fluid stress-energy tensor to describe the matter distribution, which is described by the following form
\begin{equation}\label{eq12}
T^{\mu}_{\nu}=(\rho+p_t)u^\mu u_\nu+ p_t \delta^{\mu}_{\nu}-(p_{t}-p_{r}) X^{\mu}X_{\nu},
\end{equation}
where $u^\mu$ is the fluid 4-velocity, $X_{\mu}$ is the unit space like vector in the direction of radial vector and $g_{\mu\nu}$ is the metric tensor. Furthermore, $\rho = \rho(r)$ is the energy density,  $p_r = p_r(r)$ and $p_t = p_t (r)$ are the radial and tangential pressures, respectively. The energy-momentum tensor of electromagnetic field defined by 
 \begin{equation}
 {{T^{\rm em}}^{\mu}}_{\nu}=\frac{1}{4}\big(-F^{\mu\gamma}F_{\nu\gamma}+\frac{1}{4}\delta^{\mu}_{\nu}F^{\beta\zeta}F_{\beta\zeta}\big)\,.
      \label{tem}
 \end{equation}
We define an appropriate frame of the fluid velocity vectors \cite{Cadoni:2020izk}
\beq
u^{a}=  e^{-\Phi}\delta^{a}_{0}\,, \quad X^{a}= \sqrt{1-\frac{b(r)}{r}} \delta^{a}_{1}\,,
\eeq
so that $u_{a}u^{a}=-1$ and $X_{a}X^{a}=1$ as required. Now, in its covariant form, we construct the geometrical Einstein tensor on the left-hand side with an effective energy-momentum tensor on the right-hand side, as
\begin{equation}
     R^{\mu}_{\nu}-\frac{1}{2}\delta^{\mu}_{\nu}R=8\pi \Big(T^{\mu}_{\nu}+{{T^{\rm em}}^{\mu}}_{\nu}\Big),
     \label{eq13b}
 \end{equation}
  where
 \beq
 {{T^{\rm em}}^{\mu}}_{\nu}&=&(\rho^{\rm em},p^{\rm em}_{r},p^{\rm em}_{t},p^{\rm em}_{t})\nonumber\\&=&\frac{q^{2}}{8\pi r^{4}}{\rm diag.}(-1,-1,1,1)\,.
\label{tem}
 \eeq
We start by considering a static, spherically symmetric and asymptotically flat space-time with the metric
\begin{eqnarray}\label{met}
    ds^2_{\text{stat}} = -e^{2 \Phi} dt^2 + \frac{dr^2}{1- \frac{b}{r}} + r^2 (d \theta^2 + \sin^2 \theta d \phi^2),
\end{eqnarray}
where $\Phi$ and $b$ are arbitrary functions of the variable `$r$,' where $\Phi$ is referred to as the `redshift function', and `$b$' is known as the `shape function. Taking into account the Eqs. (\ref{eq12}) and (\ref{tem}), the nonzero components of the field equations (\ref{eq13b}) take the form:
\begin{eqnarray}
&& \frac{b^{\prime}}{r^{2}} = 8\pi \Big(\rho+\frac{q^{2}}{8\pi r^{4}}\Big),  \label{eq14} \\
&& 2\left(1-\frac{b}{r}\right)\frac{\Phi^{\prime}}{r}-\frac{b}{r^{3}}  = 8\pi\Big(p_{r}-\frac{q^{2}}{8\pi r^{4}}\Big),  \label{eq15} \\
&& \left(1-\frac{b}{r}\right)\left[\Phi^{\prime\prime}+\Phi^{\prime 2}-\frac{b^{\prime}r-b}{2r(r-b)}\Phi^{\prime}-\frac{b^{\prime}r-b}{2r^2(r-b)} +\frac{\Phi^{\prime}}{r}\right] 
 = 8\pi\Big(p_{t}+\frac{q^{2}}{8\pi r^{4}}\Big),  \label{eq16}
\end{eqnarray}
where $\rho$ represents the effective energy density, $p_{r}$ is the effective radial pressure and $p_{t}$ is the effective tangential pressure respectively. 

\section{Wormhole Solutions}\label{sec3}
Following Ref.\cite{Garattini:2019ivd}, we can consider the plates positioned at a distance either parametrically fixed and radially varying. Thus, we consider a more realistic scenario, incorporating the effects of thermal fluctuations as a source of a traversable wormhole. To examine the dependence on the thermal Casimir effect, we consider plates positioned either at a fixed parametric distance or with radial variation.

\subsection{Constant Plates separation}
In this case, the form of the energy density has been proposed in Ref.\cite{Garattini:2024jkr}. It can be written as
\beq
\rho_{CP}=-\frac{k_{B}T}{8\pi d^{3}}\zeta(3)\,,
\eeq
a shape function can be straightforwardly computed using Eq.(\ref{eq14}) to obtain
\begin{eqnarray}
b(r)= c_{1}-\frac{\zeta(3) k_{B} r^3 T}{3 d^3}-\frac{q^2}{r}\,,
\end{eqnarray}
where $c_{1}$ is a integration constant. Taking $b(r_{0})=r_{0}$, it can be determined to yield  
\beq
c_{1}=\frac{\zeta(3) k_{B} r_{0}^3 T}{3 d^3}+\frac{q^2}{r_{0}}+r_{0}\,.
\eeq
Therefore,  the first EFE leads to the following shape function
\beq
b(r)&=&r_{0}+q^2 \left(\frac{1}{r_{0}}-\frac{1}{r}\right)\nonumber\\&&-\frac{\zeta(3) \left(r^3-r_{0}^3\right)}{6 d^4}\Bigg(\frac{T}{T_{\text{eff}}}\Bigg)\,,\label{beocp}
\eeq
with $\zeta(x)$ being the Riemann zeta function and $T_{\text{eff}}=1/(2dk_{B})$ in our convention. Setting $q=0$, the above result become that of Ref.\cite{Garattini:2024jkr}. From the shape function (\ref{beocp}), we can verify that the flare-out condition is always satisfied since
\ba
b'(r_{0})=\frac{q^2}{r_{0}^2}-\frac{\zeta(3) r_{0}^2}{2 d^4}\frac{T}{T_{\text{eff}}}<1\,.
\ea
This implies that
\ba
\frac{T}{T_{\text{eff}}}>\frac{2 d^4 (q^{2}-r^{2}_{0})}{\zeta(3) r_{0}^4}\,.\label{flar}
\ea
To obtain a TW, we need to compute the redshift function. We have to impose the EoS $p_{r}(r) = \omega \rho(r)$. From Eq.(\ref{eq15}) and Eq.(\ref{beocp}), one finds
\beq
\Phi'(r) = \frac{6 T_{\text{eff}} \left(d^4 \left(q^2 (r-2 r_{0})+r r_{0}^2\right)-d \zeta(3) k_{B} r^4 r_{0} T \omega \right)+\zeta  r r_{0} T \left(r_{0}^3-r^3\right)}{2 r (r-r_{0}) \left(\zeta(3)  r r_{0} T \left(r^2+r r_{0}+r_{0}^2\right)-6 d^4 T_{\text{eff}} \left(q^2-r r_{0}\right)\right)}\,.\label{Phi}
\eeq
Close to the throat, the r.h.s. of Eq.(\ref{Phid}) can be approximated by
 \beq\label{Phid1}
\Phi'(r) \simeq \frac{3 \left(d^4 \left(r_{0}^3-q^2 r_{0}\right)-\frac{\zeta(3) r_{0}^5 T \omega }{2 T_{\text{eff}}}\right)}{r_{0} (r-r_{0}) \left(\frac{3 \zeta(3) r_{0}^4 T}{T_{\text{eff}}}-6 d^4 \left(q^2-r_{0}^2\right)\right)},.
\eeq
Following Ref.\cite{Garattini:2019ivd} we can choose $\omega$ such that $\Phi'(r)=0$ to avoid the appearance of a horizon. This can be seen if
 \beq\label{Phid10}
\omega = \frac{2 d^4 T_{\text{eff}} (r^{2}_{0}-q^{2})}{\zeta(3)  r_{0}^4 T}\,.
\eeq
Note that when $q=0$, the above result reduces to that of Ref.\cite{Garattini:2024jkr}, i.e., $\omega=\frac{2 d^4 T_{\text{eff}}}{\zeta(3)  r_{0}^{2} T}$. With a flare-out condition Eq.(\ref{flar}), we find that 
 \beq\label{Phid10}
\omega = \frac{2 d^4 (r^{2}_{0}-q^{2})}{\zeta(3)  r_{0}^4 }\Bigg(\frac{T}{T_{\text{eff}}}\Bigg)<1\,,
\eeq
and can be negative when $r_{0}<q$. We can generalize the above results by incorporating the GUP corrected energy density. A new class of Casimir wormholes can be basically constructed. Following \cite{Samart:2021tvl}, it is straightforward to calculate the shape function of the traverasable wormholes with the electric charge underlying a so-called Generalized Uncertainty Principle (GUP). As proposed in Ref.\cite{Jusufi:2020rpw}, we consider the GUP corrected Casimir energy density in a constant plate scenario. We finally find
\begin{eqnarray}
b(r)=-\frac{\zeta(3) k_{B} r^3 T \left(d^2+\beta D_{i}\right)}{3 d^5}-\frac{q^2}{r}+c_2\,,
\end{eqnarray}
where $c_{2}$ is a integration constant, $D_{i}$s are the constants, see Ref.\cite{Jusufi:2020rpw}. Taking $b(r_{0})=r_{0}$, it can be determined to yield  
\beq
c_{2}=\frac{3 d^5 q^2+3 d^5 r_{0}^2+d^2 \zeta(3) k_{B} r_{0}^4 T+\beta  D_{i} \zeta(3) k_{B} r_{0}^4 T}{3 d^5 r_{0}}\,.
\eeq
Therefore, the following shape function is this case becomes
\beq
b(r)&=&r_{0}+q^2 \left(\frac{1}{r_{0}}-\frac{1}{r}\right)\nonumber\\&&-\frac{\zeta(3) \left(d^2+\beta D_{i}\right)\left(r^3+r_{0}^3\right)}{6 d^6}\Bigg(\frac{T}{T_{\text{eff}}}\Bigg)\,.\label{beogup2}
\eeq
Notice that the third term includes the GUP effect. It is also straightforward to show that $b(r)/r\rightarrow 0$ when $r\rightarrow \infty$. From the shape function (\ref{beogup2}), one finds that the flare-out condition is always satisfied since
\ba
b'(r_{0})=\frac{q^2}{r_{0}^2}-\frac{\zeta(3) r_{0}^2}{2 d^4}\frac{T}{T_{\text{eff}}}<1\,.
\ea
This implies that
\ba
\frac{T}{T_{\text{eff}}}>\frac{2 d^6 (q^{2}-r^{2}_{0})}{\zeta(3) r_{0}^4 \left(d^2+\beta D_{i}\right)}\,.\label{flar}
\ea
To obtain a TW, we have to determine the redshift function. We have to impose the EoS $p_{r}(r) = \omega \rho(r)$. From Eq.(\ref{eq15}), one finds
\beq
\Phi'(r) = -\frac{\frac{r r_{0} \left(3 d^5 r_{0}-\zeta(3)  k_{B} r^3 T (3 \omega +1) \left(d^2+\beta  D_{i}\right)+\zeta(3) k_{B} r_{0}^3 T \left(d^2+\beta D_{i}\right)\right)}{d^5}+3 q^2 (r-2 r_{0})}{2 r (r-r_{0}) \left(r r_{0} \left(-\frac{\zeta(3) k_{B} T \left(r^2+r r_{0}+r_{0}^2\right) \left(d^2+\beta  D_{i}\right)}{d^5}-3\right)+3 q^2\right)}\,,\label{Phi22}
\eeq
Close to the throat, the r.h.s. of Eq.(\ref{Phi22}) can be approximated by
 \beq\label{Phid12}
\Phi'(r) \simeq \frac{d^5 \left(r_{0}^2-q^2\right)-\zeta(3)  k_{B} r_{0}^4 T \omega  \left(d^2+\beta D_{i}\right)}{2 (r-r_{0}) \left(d^5 \left(r_{0}^2-q^2\right)+\zeta(3)  k_{B} r_{0}^4 T \left(d^2+\beta D_{i}\right)\right)}\,.
\eeq
Following Ref.\cite{Garattini:2019ivd}, we can choose $\omega$ such that $\Phi'(r)=0$ to avoid the appearance of a horizon. This can be done if
 \beq\label{Phid1}
\omega \simeq \frac{d^5 \left(r_{0}^2-q^2\right)}{\zeta(3)  k_{B} r_{0}^4 T \left(d^2+\beta D_{i}\right)}\equiv \frac{2d^6  \left(r_{0}^2-q^2\right)}{\zeta(3) r_{0}^4 \left(d^2+\beta D_{i}\right)}\Bigg(\frac{T_{\text{eff}}}{T}\Bigg)\,.
\eeq
The result is more general compared to the previous case. Note that when $q=0,\,\beta=0$, the above result reduces to that of Ref.\cite{Garattini:2024jkr}, i.e., $\omega=d^3/(\zeta(3) k_{B} r_{0}^2 T)$.

\subsection{Variable Plates separation}
In the same manner of previous work on zero temperature Casimir wormholes \cite{Garattini:2019ivd}, the plates separation distance $d$ can be promoted to a radial variable $r$. We can compute the form of the energy density. For the high temperature case, the thermal energy density is given by \cite{Garattini:2024jkr}
\beq
\rho_{VP}=-\frac{k_{B}T}{8\pi r^{3}}\zeta(3)\,,
\eeq
where for in this case, we also have that the EoS leads to $\omega_{TC}=2$. Following Ref.\cite{Garattini:2024jkr} for high temperature regimes, we come up with the following shape function:
\begin{eqnarray}
b(r)= -\zeta(3) k_{B} T \log (r)-\frac{q^2}{r}+c_1\,,
\end{eqnarray}
where $c_{1}$ is a integration constant. Taking $b(r_{0})=r_{0}$, it can be determined to yield  
\beq
c_{1}=\frac{\zeta(3) k_{B} T r_{0}\log (r_{0})+q^2+r_{0}^2}{r_{0}}\,.
\eeq
Therefore,  the first EFE leads to the following shape function
\beq
b(r)&=&r_{0}+q^2 \left(\frac{1}{r_{0}}-\frac{1}{r}\right)\nonumber\\&&-\zeta(3)k_{B}T \log \left(\frac{r}{r_{0}}\right)\,,\label{beo}
\eeq
with $\zeta(x)$ being the Riemann zeta function. Having considered the charge contribution, the above result is a generalization of what is found in Ref.\cite{Garattini:2024jkr}. We also find that $b(r)/r\rightarrow 0$ when $r\rightarrow \infty$. From the shape function (\ref{beo}), one finds that the flare-out condition is always satisfied since
\beq
b'(r_{0})=\frac{q^2}{r_{0}^2}-\frac{\zeta(3)k_{B}T}{r_{0}}<1\,,\label{beof}
\eeq
implying that 
\beq
q^{2} -r^{2}_{0}< \zeta(3)k_{B}T r_{0}\,.\label{beoq}
\eeq
Moreover from the condition $1-b(r)/r>0$, it is possible to clarify that there exists $r=r^{*}$ such that $b(r^{*})=0$, where
\beq
r^{*} = -\frac{q^2}{\zeta(3) k_B T\,W\left(-\frac{q^2 e^{-\frac{(q^2+r_0^2)}{\zeta(3)  r_0 T k_B}}}{\zeta(3)  r_0 T k_B}\right)}=r_{0}e^{\frac{(\varepsilon+1)r_0}{\zeta(3) T k_B}}e^{W(\chi)}\,,\label{rstar}
\eeq
with 
\beq
\chi\equiv -\frac{r_{0}\varepsilon e^{-\frac{(\varepsilon+1)r_0}{\zeta(3) T k_B}}}{\zeta(3) T k_B}\quad{\rm and}\quad \varepsilon\equiv \frac{q^{2}}{r^{2}_{0}}\,,
\eeq
and $W(\chi)$ is a Lambert $W$ function or product logarithm and we have used the identity $e^{W(x)}=\tfrac{x}{W(x)}$. From the above result, Let us focus on a small charge approximation, $\varepsilon=q^{2}/r^{2}_{0}\ll 1$. In this regime, we find
\beq
r^{*} \simeq r_0 e^{\frac{r_0}{\zeta(3)  T k_B}}+\frac{q^2 \left(e^{\frac{r_0}{\zeta(3)  T k_B}}-1\right)}{\zeta(3)  T k_B}+{\cal O}\left(q^3\right)\,.
\eeq
The above result shows that the point $r^{*}$ where $b(r^{*})=0$ is positive and greater than $r_0$. Notice that at the zeroth order, it was already found in Ref.\cite{Garattini:2024jkr}. The second term emerges thank to the charge contribution. To obtain a TW we need to compute the redshift function. We have to impose the EoS $p_{r}(r) = \omega \rho(r)$. From Eq.(7), one finds
\beq
\Phi'(r) = \frac{-b(r) r+\zeta(3) k_B T r \omega +q^2}{2 r^2 (b(r)-r)}\,.\label{Phi}
\eeq
Substituting $b(r)$ from Eq.(\ref{beo}), we come up with
\beq\label{Phid}
\Phi'(r) =\frac{r r_0 \left(r_0-\zeta(3)  T \omega  k_B\right)-\zeta  r r_0 T k_B \log \left(\frac{r}{r_0}\right)+q^2 \left(r-2 r_0\right)}{2 r \left(\zeta  r r_0 T k_B \log \left(\frac{r}{r_0}\right)+\left(r-r_0\right) \left(r r_0-q^2\right)\right)}\,.
\eeq
 Close to the throat, the r.h.s. of Eq.(\ref{Phid}) can be approximated by
 \beq\label{Phid1}
\Phi'(r) \simeq \frac{r^{2}_0 \left(r_0-\zeta(3)  T \omega  k_B\right)-q^2 r_0}{2 r_{0}\left(r-r_0\right) \left(r^{2}_0-q^2\right)}\,.
\eeq
Following Ref.\cite{Garattini:2019ivd} we can choose $\omega$ such that $\Phi'(r)=0$ to avoid the appearance of a horizon. This can be seen if
 \beq\label{Phid10}
\omega = \frac{r_0 \left(r^2_{0}-q^2\right)}{\zeta(3)  r^2_{0} T k_B}\,.
\eeq
Note that when $q=0$, the above result reduces to that of Ref.\cite{Garattini:2024jkr}, i.e., $\omega=r_{0}/(\zeta(3)  r^2_{0} T k_B)$. However, we can separately consider for the two cases:
\beq\label{Phid101}
\omega = \frac{r_0 \left(1-\varepsilon\right)}{\zeta(3) T k_B} \simeq
\begin{cases}
\frac{r_{0}}{\zeta(3) k_{B} T}-\frac{r_{0}\varepsilon}{\zeta(3) k_{B} T}\quad{\rm for}\quad \varepsilon\ll 1,\\-\frac{r_{0}\varepsilon}{\zeta(3) T k_B}\quad\quad\quad\quad\,\,{\rm for}\quad \varepsilon\gg 1\,.
\end{cases}
\eeq
Interestingly, this allow us to compute the throat of the wormhole: $r_{0}$. Here using Eq.(\ref{Phid10}) and incorporating with $\omega_{VP}=2$, we simply find a positive solution:
\beq
r_{0}=\zeta(3)  T k_B\Big(1+ \sqrt{1+ \frac{q^2}{\zeta(3)^2 T^2 k_B^2}}\Big)\,.
\eeq
Having compared with Ref.\cite{Garattini:2024jkr}, in our convention, we take $l_{P}=1$. Notice from the above result that the size of the throat is enlarged further by the effect of the charge. Moreover, it can be more generalized by following the work present in Ref.\cite{Jusufi:2020rpw} for elaborating in more details about the GUP corrected energy density. This allow to construct a new class of Casimir wormholes.
Following \cite{Samart:2021tvl}, it is straightforward to calculate the shape function of the traverasable wormholes with the electric charge underlying a so-called Generalized Uncertainty Principle (GUP). We finally find
\begin{eqnarray}
b(r)=\frac{\beta D_{i} \zeta(3) k_{B} T}{2 r^2}-\zeta(3) k_{B} T \log (r)-\frac{q^2}{r}+c_1\,,
\end{eqnarray}
where $c_{1}$ is a integration constant, and $D_{i}$s are the constants, see Ref.\cite{Jusufi:2020rpw}. Taking $b(r_{0})=r_{0}$, it can be determined to yield  
\beq
c_{1}=-\frac{\beta D_{i} T \zeta(3) k_{B}}{2 r_{0}^2}+T \zeta(3) k_{B} \log (r_{0})+\frac{q^2}{r_{0}}+r_{0}\,.
\eeq
Therefore,  the first Einstein Field Equation leads to the following shape function
\beq
b(r)&=&r_{0}+q^2 \left(\frac{1}{r_0}-\frac{1}{r}\right)+T \zeta(3) k_{B} \left[\log\bigg(\frac{r}{r_{0}}\bigg)+\frac{\beta  D_{i}}{2} \left(\frac{1}{r^2}-\frac{1}{r_0^2}\right)\right]\,.\label{beogup}
\eeq
Notice that the last term is due to the GUP effect, while the last term is the consequences of thermal fluctuations. It is also straightforward to show that $b(r)/r\rightarrow 0$ when $r\rightarrow \infty$. From the shape function (\ref{beogup}), one finds that the flare-out condition is always satisfied since
\beq
b'(r_{0})=\frac{q^2}{r_0^2}-T \zeta(3) k_{B} \left(\frac{\beta  D_i}{r_0^3}+\frac{1}{r_0}\right)<1\,,\label{beof}
\eeq
implying that 
\beq
q^{2}-r^{2}_{0}< T \zeta(3) k_{B} \left(r_{0}+\frac{\beta  D_i}{r_0}\right)\,.\label{beoq}
\eeq

To obtain a TW, we are going to compute the redshift function. We have to impose the EoS $p_{r}(r) = \omega \rho(r)$. From Eq.(7), one finds
\beq
\Phi'(r) = \frac{\beta  T \chi_B D_i \left(r_0^2 (1-2 \omega )-r^2\right)+2 r r_0 \left(-r_0 \left(r T \omega  \chi_B+r T \chi_B \log \left(\frac{r}{r_0}\right)+2 q^2\right)+q^2 r+r r_0^2\right)}{2 r \left(\beta  \left(r^2-r_0^2\right) T \chi_B D_i+2 r r_0 \left(r_0 \left(r T \chi_B \log \left(\frac{r}{r_0}\right)+q^2+r^2-r r_0\right)-q^2 r\right)\right)}\,,\label{Phi}
\eeq
with $\chi_{B}\equiv \zeta(3) k_{B}$. Close to the throat, the r.h.s. of Eq.(\ref{Phi}) can be approximated by
 \beq\label{Phid12}
\Phi'(r) \simeq \frac{-T \zeta(3) k_{B}  \omega  \left(\beta  D_{i}+r_{0}^2\right)-q^2 r_{0}+r_{0}^3}{2 (r-r_{0}) \left(\beta D_{i} T \zeta(3) k_{B} -q^2 r_{0}+r_{0}^3\right)}\,.
\eeq
Following Ref.\cite{Garattini:2019ivd}, we can choose $\omega$ such that $\Phi'(r)=0$ to avoid the appearance of a horizon. This can be done if
 \beq\label{Phid1}
\omega = -\frac{q^2 r_{0}-r_{0}^3}{\beta D_{i} T \zeta(3) k_{B} +r_{0}^2 T \zeta(3) k_{B}}\,.
\eeq
The result is more general compared to the previous case. Note that when $q=0,\,\beta=0$, the above result reduces to that of Ref.\cite{Garattini:2024jkr}, i.e., $\omega=1/(r_{0} T \zeta(3) k_{B})$. Following the preceding result, we can also compute the throat of the wormhole. In case of a small $\beta\ll 1$, we can expand $\omega$ to obtain
\beq
\omega \approx \frac{\left(r_{0}^2-q^2\right)}{r_{0} T \zeta(3) k_{B}}\left(1+\frac{\beta D_{i}}{r_{0}^2}\right)\,.
\eeq
Notice that the charge contribution as well as the GUP correction appear in the above expression.

\section{Concluding Remarks}
In this work, following Ref.\cite{Garattini:2024jkr}, we have investigated the consequences of thermal fluctuations to the Casimir effect on a traversable wormhole subject to the charge contribution. To perform the calculations, we focused on scenarios where the plates are constant and radially varying. For the high temperature approximation, we first considered only charge contribution to the thermal Casimir wormholes. We investigated the effect of such finite temperature effects incorporating with a charge on the size of a wormhole. Having compared with Ref.\cite{Garattini:2024jkr}, the size of the throat is enlarged further by the effect of the charge contribution in the case when plates are radially varying given by
\beq
r_{0}=\zeta(3)  T k_B\Big(1+ \sqrt{1+ \frac{q^2}{\zeta(3)^2 T^2 k_B^2}}\Big)\,.
\eeq
We have verified that the flare-out condition is always satisfied. Indeed, we can restore constants, e.g, $G,\,c,\,\hbar$, so that our results reflect physical interpretations. This is done by introducing the Casimir thermal wave length Ref.\cite{Garattini:2024jkr}. The second scenario deals with the GUP corrections to the thermal Casimir energy. In this case, we have also found a new class of WH solutions. We have found that the flare-out condition is always satisfied in this case. Interestingly, the wormhole solutions in the present work can be further tested by the wormhole observations, see, e.g., \cite{Abe:2010ap,Toki:2011zu,Godani:2021aub,Ohgami:2015nra,Dai:2019mse}.

\acknowledgments
This work is financially supported by Thailand NSRF via PMU-B under grant number PCB37G6600138.

\end{document}